\documentclass[twocolumn,showpacs,prb,superscriptaddress]{revtex4}
\usepackage{graphicx}
\usepackage{dcolumn}
\usepackage{amsmath}
\usepackage{amssymb}

\makeatletter
\def\btt#1{\texttt{\@backslashchar#1}}
\DeclareRobustCommand\bblash{\btt{\@backslashchar}} \makeatother

\begin{document}

\title{Majorana Fermions and Odd-frequency Cooper Pairs in a Nano Wire}

\author{Yasuhiro Asano}
\affiliation{Department of Applied Physics and Center for Topological Science \& Technology,
Hokkaido University, Sapporo 060-8628, Japan}

\author{Yukio Tanaka}
\affiliation{Department of Applied Physics, Nagoya University, Nagoya 464-8603, Japan}

\date{\today}

\begin{abstract}
We discuss a strong relationship between Majorana fermions and 
odd-frequency Cooper pairs which appear at 
a disordered normal (N) nano wire attached to a topologically 
nontrivial superconducting (S) one. 
The transport properties in superconducting nano wire junctions show 
universal behaviors irrespective of the degree of disorder: 
the quantized zero-bias differential conductance at $2e^{2}/h$ in NS junction 
and the fractional current-phase ($J-\varphi$) relationship of the 
Josephson effect in SNS junction $J\propto \sin(\varphi/2)$. 
Such behaviors are exactly the same as those found in the anomalous 
proximity effect of odd-parity spin-triplet superconductors.
We show that the odd-frequency pairs exist wherever the Majorana fermions stay.
\end{abstract}


\maketitle

Majorana fermion (MF) satisfying a special relation of $\gamma=\gamma^{\dagger}$with $\gamma$ ($\gamma^{\dagger}$) being the annihilation (creation) operator
 has been an exciting object
since the original prediction by Majorana~\cite{Majorana}.  
Finding of MFs and controlling of Majorana bound states (MBSs) 
are hot research issues
in condensed matter physics~\cite{Majorana1}
from the view of potential application of MBS to the topological 
quantum computation~\cite{Ivanov,Nayak}. 
To date, we have known several promising systems hosting MFs 
such as spin-triplet 
$p$-wave superconductors~\cite{Read,Bolech,Sengupta,Kitaev}, 
topological insulator /superconductor heterostructures~\cite{Fu2008}%,Fu2009,Law}, 
semiconductor / superconductor junctions with strong spin-orbit coupling
~\cite{Sato,DasSarma2010,Alicea,Potter}, helical superconductors \cite{helical}, 
and superconducting topological insulators \cite{STI1}. 
Most attracting case among them is the semiconductor nano wire 
fabricated on top of a superconductor because of
its easy tunability of MBS 
by changing the chemical potential in the nano wire
and by applying the Zeeman field onto 
it~\cite{Lutchyn2010,Oreg}. 
Actually, a plenty of 
theoretical studies have discussed MFs or MBSs 
in such nano wires~\cite{Tewari,Golub}. 
The zero-bias conductance peak reported 
%by Kouwenhoven's' group 
very recently would be considered as an evidence of MFs (MBS)
~\cite{Lutchyn2010,Oreg,Delft1}. 
These researches have stimulated a number of theoretical 
investigation on unusual charge transport phenomena through the MBS in 
normal-metal / superconductor (NS) and 
superconductor /normal-metal/ superconductor (SNS) junctions
on nano 
wires~\cite{Akhmernov}. 
However, no studies have ever tried to analyze features of Cooper pairs 
which support the anomalous transport properties. 
We address this issue in the present Letter.

Odd-frequency Cooper pairing was originally proposed to understand nature of
unconventional superfluidity and 
superconductivity\cite{berezinskii}.
Ubiquitous appearance of the odd-frequency pairs 
at the surface of superconductors and near the interface 
of superconducting junctions 
has been established and widely accepted in recent years~\cite{bergeret,tanaka07e}. 
The zero-energy Andreev bound state (ABS) 
at the surface of unconventional superconductors
~\cite{ABS1}
is reinterpreted in terms of the odd-frequency Cooper 
pairing~\cite{tanakareview}. 
In particular, the odd-frequency Cooper pairs make the background of 
the anomalous proximity effect in a diffusive normal metal attached to 
a spin-triplet superconductor~\cite{tanaka07}: 
(i) the large zero-energy quasiparticle density of states 
in a normal metal~\cite{yt05r,yt04}, (ii) the quantized zero-bias conductance at twice of the Sharvin's value 
in diffusive NS junctions~\cite{yt04}, (iii) the fractional current($J$)-phase($\varphi$) 
relationship of 
$J\propto \sin(\varphi/2)$ in diffusive SNS junctions~\cite{ya06}, 
(iv) the zero-bias anomaly in nonlocal conductance spectra~\cite{ya07}
and (v) the anomalous surface impedance in NS bilayers~\cite{ya11}.

In this Letter, we show that disordered NS and SNS junctions of nano wire 
indicate the properties of (i)-(iii) when the superconducting nano wire is 
topologically 
nontrivial.
In addition, the amplitude of odd-frequency pairs in the normal nano wire suddenly 
grows 
as soon as the superconducting nano wire undergoes the transition to 
topologically nontrivial phase.
The unusual transport phenomena due to the MFs~\cite{Delft1} are nothing but 
the anomalous proximity effect due to the odd-frequency pairs.
We will conclude that the odd-frequency Cooper pairs are indispensable to 
realizing MFs in solids.

\begin{figure}[tbh]
\begin{center}
\includegraphics[width=7cm]{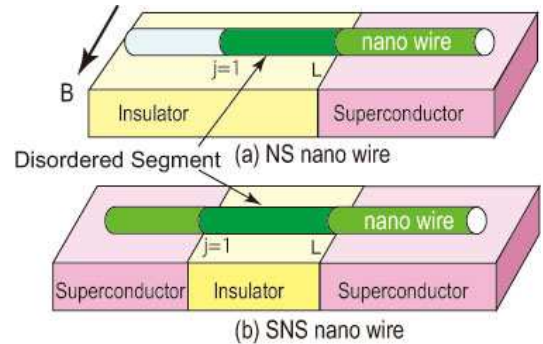}
\end{center}
\caption{(color online). 
Schematic pictures of NS and SNS junctions. 
}
\label{fig1}
\end{figure}

Let us consider a nano wire with strong spin-orbit coupling fabricated 
on a junction of an insulator and a metallic superconductor 
as shown in Fig.~\ref{fig1}. 
A segment on the insulator and 
that on the superconductor are in the normal
and the superconducting states, respectively. 
The diameter of the nano wire is sufficiently 
small so that the number of propagating channel is unity for each spin 
degree of freedom. We describe the present 
nano wire by using the tight-binding model in one-dimension, 
for noninteracting electrons~\cite{interaction},
\begin{align}
H_0=& -t \sum_{j,\alpha} \left( c_{j+1,\alpha}^\dagger c_{j,\alpha} + c_{j,\alpha}^\dagger c_{j+1,\alpha}
\right) \nonumber\\
+i&\frac{\lambda}{2} \sum_{j,\alpha,\beta}
\left[ c_{j+1,\alpha}^\dagger (\hat{\sigma}_2)_{\alpha,\beta}c_{j,\beta}  
- c_{j,\alpha}^\dagger (\hat{\sigma}_2)_{\alpha,\beta} c_{j+1,\beta}
\right] \nonumber\\
+& \sum_{j,\alpha,\beta} c_{j,\alpha}^\dagger\left\{ (2t-\mu) \hat{\sigma}_0 - V_{ex} \hat{\sigma}_3\right\}_{\alpha,\beta} c_{j,\beta}, \label{h0} \\
H_s=& \sum_{ j\geq L+1} \left[ \Delta e^{i \varphi }c_{j,\uparrow}^\dagger c_{j,\downarrow}^\dagger + \text{H.c.}\right],
\label{hdelta}\\
H_d=& \sum_{1\leq j\leq L,\alpha} V_{j} c_{j,\alpha}^\dagger c_{j,\alpha},\\
H_{s2}=& \sum_{ j\leq 0} \left[ \Delta e^{i \varphi_2 }c_{j,\uparrow}^\dagger c_{j,\downarrow}^\dagger + \text{H.c.}\right],
\end{align}
where $c_{j,\alpha}^\dagger (c_{j,\alpha})$ is the creation (annihilation) operator of an electron 
at the lattice site $j$ with spin $\alpha=(\uparrow$ or $\downarrow$), 
$t$ denotes the hopping integral, $\mu$ is the chemical potential, 
and $\Delta$ is the pair potential in the superconducting segment.
The Pauli matrices in spin space are denoted by $\hat{\sigma}_j$ for $j=1-3$ and the unit matrix of $2\times 2$ 
is $\hat{\sigma}_0$.
The on-site potential in the normal segment is given randomly in the range of $-W/2 \leq V_j \leq W/2$.
We measure the energy and the length in units of $t$ and the lattice constant, respectively.
Throughout this paper, we fix several parameters as $\mu=t$, $W=2t$, and the pair potential 
at the zero temperature $\Delta=0.01t$. 
The number of samples used for the random ensemble averaging is typically $10^3-10^5$.
By tuning the magnetic field $B$
as shown in Fig.~\ref{fig1}, it is possible to 
introduce external Zeeman potential $V_{ex}$.
For $V_{ex}>V_c \equiv \sqrt{\Delta_0^2+\mu^2}$, 
the number of propagating channels becomes unity and
the superconducting segment undergoes the transition to topologically 
nontrivial phase.
In the tight-binding model, the finite band width gives an additional 
condition $V_{ex}<V_{c2}\equiv 4t-\mu$ for the topological phase. 
%Since the normal segment is in the localization regime, 
%two segments of nano wire are weakly coupled and 
%MF (MBS) appears at the boundary between two segments. 
%In the following, we mainly study the transport properties of two nano wires.
%One is the nano wire with $V_{ex}=1.5t$ and $\lambda=0.5t$, which we call 
%topological nano wire hosting MF. 
%The other is the non-topological 
%nano wire with $V_{ex}=\lambda=0$ in which MF is absent. 
%The latter one is studied as a reference.
%By comparing the results of the two nano wires, we discuss anomalous 
%charge transport properties of the topological nano wire.
Here we briefly summarize calculated results of the normal conductance of 
the disordered nano wires with using the recursive Green function method~\cite{Lee}. 
By analyzing the Hamiltonian $H_0+H_d$, 
we confirmed that the normal 
conductance decays exponentially with increasing $L$~\cite{localization}.
This is because one-dimensional disordered wires are basically in the localization regime. 
\begin{figure}[tbh]
\begin{center}
\includegraphics[width=8.5cm]{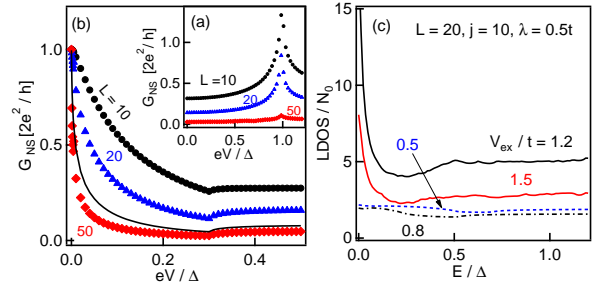}
\end{center}
\caption{(color online).
The differential conductance of NS nano wires is plotted as a function of the bias-voltage for 
non-topological nano wire ($V_{ex}=\lambda=0$) in (a) and for the topological nano wire 
( $V_{ex}=1.5t$ and $\lambda=0.5t$) in (b). In (b), we also plot the results for $W/t=4$ and $L=10$ 
with a solid line.
The local density of states at the center of the disordered segment 
($j=10$) are shown for several $V_{ex}$ in (c) with $\lambda=0.5t$, where 
we add a small imaginary part $i\delta_\epsilon$ with $\delta_\epsilon=0.01\Delta$ to the energy.
}
\label{fig2}
\end{figure}

At first, we calculate 
the differential conductance $G_{\textrm{NS}}$ of NS junctions based on the 
standard formula ~\cite{BTK}, 
\begin{align}
G_{\textrm{NS}} = \frac{e^2}{h} \sum_{\alpha,\beta}
\left[ \delta_{\alpha,\beta} - |r^{ee}_{\alpha,\beta}|^2+|r^{he}_{\alpha,\beta}|^2 \right]_{E=eV},
\label{gns}
\end{align}
where we consider the Hamiltonian 
$H_0+H_d+H_s$.
In Eq.~(\ref{gns}), $r^{ee}_{\alpha,\beta}$ and $r^{he}_{\alpha,\beta}$ are the normal and Andreev 
reflection 
coefficients of the junction at energy $E$ measured from the fermi level.
We show $G_{\textrm{NS}}$ in units of $G_Q=2e^{2}/h$
as a function of the bias-voltage $eV$ for the non-topological and the topological nano wires 
in Figs.~\ref{fig2}(a) and (b), respectively. The length of disordered segment $L$ is chosen as
10, 20, and 50 lattice constants.
The conductance for the non-topological nano wires ( $V_{ex}=\lambda=0$ ) in (a) decreases with 
increasing the length 
of disordered segment $L$ for all $eV$. 
The similar tendency can be seen also in the 
conductance of the topological nano wires ( $V_{ex}=1.5t$ and $\lambda=0.5t$) in (b) for finite $eV$. 
However the zero-bias conductance
of the topological nano wires is quantized at $G_Q$ irrespective of $L$, 
which is an intrinsic phenomenon in the presence of MF.%~\cite{Flensberg,Qu,Wimmer}. 
The results suggest a perfect transmission channel 
due to the penetration of a resonant state into the disordered segment.
The local density of states (LDOS) in the disordered nano wire supports this statement as
shown in Fig.~\ref{fig2}(c), where we plot the LDOS at the center of the disordered 
segment ($j=10$) as a function of $E$ for $L=20$ and $\lambda=0.5t$. 
The results are normalized to the density of states at 
the fermi level in clean normal nano wire $N_0$.
The LDOS for $V_{ex}>V_{c}$ show the large zero-energy peak reflecting the MBS 
as shown in the results for $V_{ex}/t=1.5$ and 1.2.
On the other hand, the LDOS for $V_{ex}/t=0.5$ and 0.8 are almost flat around the zero energy.

\begin{figure}[tbh]
\begin{center}
\includegraphics[width=8.5cm]{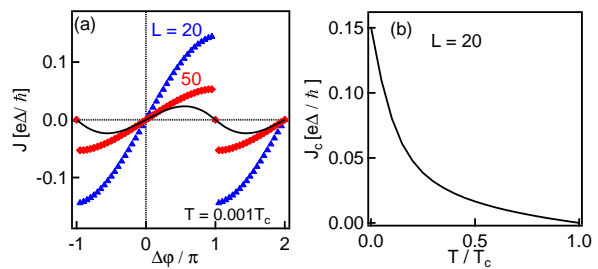}
\end{center}
\caption{(color online). (a): Current-phase relationship in SNS junctions 
of topological wire at $T/T_c=0.001$ for $V_{ex}=1.5t$ and $\lambda=0.5t$.
For comparison, the results for non topological wire ($V_{ex}=\lambda=0$ and $L=50$) 
is plotted with a solid line. 
(b): The Josephson critical current versus temperature in a topological nano wire. 
%The dependence of the pair potential on temperature is calculated 
%based on the BCS theory.
}
\label{fig3}
\end{figure}

Secondly, we explain why the superconducting nano wire shows the anomalous proximity effect 
which is unique to the $p_x$-wave spin-triplet superconductor.
The single particle Hamiltonian in Eq.~(\ref{h0}) is essentially equivalent to 
$\hat{h}_0(k) = \xi_k \hat{\sigma}_0 - \lambda k \hat{\sigma}_2 -V_{ex} \hat{\sigma}_3$ 
in momentum space with $\xi_k=\hbar^2k^2/(2m)-\mu$. 
By applying a unitary trasformation diagonalizing $h_0(k)$ and $-h_0^\ast(-k)$, 
Bogoliubov-de Gennes (BdG) Hamiltonian of the nanowire reduces to $2\times 2$ Hamiltonian
for $V_{ex}>V_c$ (See also Appendix A for details),
 \begin{align}
\left[ \begin{array}{cc}
\hat{h}_0(k) & i\Delta \hat{\sigma}_2\\
-i\Delta^\ast \hat{\sigma}_2 & -\hat{h}_0^\ast(-k)
\end{array}\right]
\to
\left[ \begin{array}{cc}
\xi_k-A & \tilde{\Delta}_k e^{i\pi/2}\\
\tilde{\Delta}_k e^{-i\pi/2}& -\xi_k+A
\end{array}\right],\label{unitary}
\end{align}
with $A=\sqrt{V_{ex}^2+(\lambda k)^2}$ and $\tilde{\Delta}_k=\Delta\lambda k/A$ 
because a spin branch pinches off from the 
fermi level (i.e., $\xi+A>0$).
The right-hand side of Eq.~(\ref{unitary}) is equivalent to the BdG Hamiltonian 
of spin less $p_x$-wave superconductor in one-dimension
when we focus on low energy excitation after
redefining the chemical potential $\mu+A \to \mu$ and the pair potential
$\tilde{\Delta}_k e^{i\pi/2}\to \Delta (k/k_F)$ with $k_F$ being the fermi wave number. 
Therefore physics in the topological nanowire is the same as that of $p_x$-wave superconductor.
In fact, we have confirmed that the Josephson current in SNS junctions of disordered nanowire
show the fractional current-phase relationship 
at low temperature as shown in Fig.~\ref{fig3}(a)~\cite{ya06}. Here 
 we attach the second supercondutor for $j\leq 0$ by adding $H_{s2}$ to $H_0+H_s+H_d$, 
and plot the Josephson current $J$ as a function of $\Delta\varphi=\varphi_2-\varphi$ at $T=0.001T_c$
for $V_{ex}=1.5t$ and $\lambda=0.5t$.
The results show $J\propto \sin(\Delta\varphi/2)$ for $-\pi\leq \Delta\varphi\leq\pi$ irrespective of $L$. 
For comparison, we also plot the results for $V_{ex}=\lambda=0$ and $L=50$ with a solid line which shows 
usual sinusoidal current phase relationship. Correspondingly the Josephson critical current plotted as a function 
of temperature shows so called low temperature anomaly~\cite{ya06} in Fig.~\ref{fig3}(b).

Next we discuss the relationship between Majorana fermions and odd-frequency Cooper pairs
by analyzing the Green functions in junctions of $p_x$-wave supercondutor.
A semi-infinite wire of 
$p_x$-wave superconductor occupying $x>0$ hosts a Majorana fermion around its edge at $E=0$. 
Solving the BdG equation, the wave function of such surface state is 
calculated to be $\phi_0(x)^T=[u_0(x),v_0(x)]^T$,
 where $u_0(x)=C(x) \chi $, $v_0(x)=C(x)\chi^\ast$,    
$C(x)=\sqrt{2/\xi_0}e^{-x/2\xi_0}\sin(kx)$, $\chi= e^{i\pi/4}e^{i\varphi/2}$, and
$\xi_0$ is the coherence length.
The electron operator includes the contribution 
from such surface state $\psi_0(x)$ as represented by
%\begin{align}
$\psi_0(x) = \chi \gamma(x)$, $ \psi_0^\dagger(x) = \chi^\ast \gamma(x)$
%, 
%\label{psi0}
%\end{align} 
with $\gamma(x) =C(x) (\gamma_0+\gamma_0^\dagger)$. 
Here 
%$\psi_0(x)$ is the annihilation operator of an electron corresponding surface 
%state, 
$\gamma_0$ is the annihilation operator of the Majorana bound state. 
The special relation $v_0(x)=u_0^\ast(x)$ plays a crucial role in 
the Majorana relation of $\gamma(x)=\gamma^\dagger(x)$~\cite{tanakareview}. 
As a result, the two Green 
functions calculated for $|E|\ll \Delta$ 
%\begin{align}
%G(x,t;x',t')=&-i\Theta(t-t')\left\langle \left\{ \psi(x,t), \psi^\dagger(x,t) \right\} \right\rangle,\\
%F(x,t;x',t')=&-i\Theta(t-t')\left\langle \left\{ \psi(x,t), \psi(x,t) \right\} \right\rangle.
%\end{align} 
\begin{align}
g(E;x,x') &\approx \frac{u_0(x)u_0^\ast(x')+v_0^\ast(x)v_0(x')}{E+i\delta_\epsilon},\\
f(E;x,x') &\approx \frac{u_0(x)v_0^\ast(x')+v_0^\ast(x)u_0(x')}{E+i\delta_\epsilon},
\end{align} 
depend on each other. Since $v_0(x)=u_0^\ast(x)$,
they satisfy 
\begin{align}
g(E,x,x') = (\chi^\ast)^2f(E,x,x') =I(E;x,x').\label{gf}
\end{align}
This relation directly 
%comes from the Majorana relation in Eq.~(\ref{psi0}) and 
bridges Majorana fermions and odd-frequency Cooper pairs.
The real (imaginary) part of $(\chi^\ast)^2f(E;x,x)$ is 
an odd (even) function of $E$, which represents the odd-frequency symmetry of Cooper pairs.
The orbital part is $s$-wave symmetry when $f$ is calculated at $x=x'$.
In fact, the imaginary part of $g(E;x,x)$ must be even function of $E$ because it represents 
LDOS of the Majorana bound state.
It is possible to check this argument in a junction which consists of a 
normal metal $(x<0)$ and a $p_x$-wave superconductor ($x>0$) in one-dimension\cite{ya04-2} 
(See also Appendix B).
At the NS interface ($x=0$), we introduce a potential barrier $V_0\delta(x)$ whose normal 
transmission coefficient is $t_n=k_F/(k_F+iz_0)$ with $z_0=V_0/\hbar v_F$.
When we focus on the subgap energy $|E|\ll \Delta$ in the tunneling limit $|t_n|\ll 1$, we 
find that the Green functions in the superconductor $x>0$ satisfy Eq.~(\ref{gf}) and become 
\begin{align}
I(E;x,x)
\approx 
\frac{\pi N_0 \Delta}{E+i\Delta|t_n|^2/2}e^{-x/\xi_0} \sin^2(kx). \label{gf_ns}
\end{align}
For comparison, we show the anomalous Green function in a uniform $p_x$-wave superconductor
\begin{align}
(\chi^\ast)^2f(E;x,x')=& -i\frac{\pi N_0}{2} \frac{\Delta\sin k(x-x')}{\sqrt{(E+i0^+)^2-\Delta^2}},\label{funiform}
\end{align}
with $\sqrt{(E+i0^+)^2-\Delta^2}$ being $\textrm{sgn}(E)\sqrt{E^2-\Delta^2}$ for $|E| > \Delta$ and
$i\sqrt{\Delta^2-E^2}$ for $|E| < \Delta$,
where we assume $|x-x'| \ll \xi_0$.
The anomalous Green function satisfies $f(x-x')=-f(x'-x)$ reflecting the odd-parity symmetry.
In contrast to Eq.~(\ref{gf_ns}), the real (imaginary) part of $(\chi^\ast)^2f(E,x,x') $ 
is an even (odd) function of $E$, which represents the even-frequency symmetry.

\begin{figure}[h]
\begin{center}
\includegraphics[width=7.0cm]{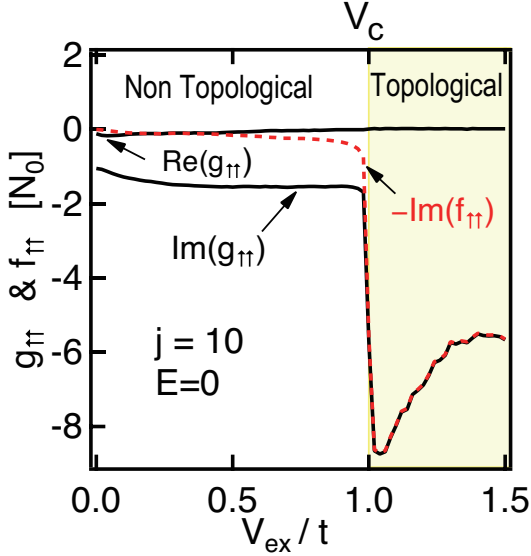}
\end{center}
\caption{(color online). 
The normal 
$g_{\upuparrows}$ and the anomalous $f_{\upuparrows}$ Green function for $E=0$ at $j=10$,
with $L=20$. We introduce a small imaginary part $\delta_\epsilon=0.001\Delta$.
%The results are normalized to the normal density of states at $V_{ex}=0$ ($N_0$).
}
\label{fig4}
\end{figure}

The important relation in Eq.~(\ref{gf}) can be confirmed in the normal segment of NS 
nano wire as shown in Fig.~\ref{fig4}, where we fix the energy at $E=0$ and plot 
$g_{\upuparrows}(j,j)$ and $-f_{\upuparrows}(j,j)$ at the center of the normal segment $j=10$.
We note that an extra phase factor $\varphi=\pi/2$ in Eq.~(\ref{unitary}) makes $(\chi^\ast)^2=-1$ 
in Eq.~(\ref{gf}).
For $V_{ex}>V_c$, the results show $\textrm{Im}(g_{\upuparrows}) = -\textrm{Im}(f_{\upuparrows})$. 
The real part of $f_{\upuparrows}$ is always zero at $E=0$ due to the odd-frequency symmetry. 
Correspondingly, $\textrm{Re}(g_{\upuparrows})$ also goes to zero for $V_{ex}>V_c$.
 In addition, $-\textrm{Im}(g_{\upuparrows})=\textrm{Im}(f_{\upuparrows})$ 
suddenly increases as $V_{ex}$ increasing across $V_c$, which corresponds to the 
zero-energy peak in LDOS in Fig.~\ref{fig2}(c). The results demonstrate the 
penetration of Majorana fermions and odd-frequency Cooper pairs 
into the normal disordered segment at the same time. 

In the NS junction of $p_x$-wave superconductor, it is possible to calculate exactly the wave function 
in the presence of a single impurity $V_i\delta(x-x_i)$ 
in the normal metal by solving the Lippmann-Schwinger equation~\cite{ya04-2} (See also Appendix C),
\begin{align}
\phi_n(x)=&\phi_n^{ini}(x) + \hat{G}(E;x,x_i) V_i \hat{\sigma}_3 \phi_n(x_i),
\label{ls}
\end{align}
where $\phi_n(x)$ is the wave function in the presence of 
the impurity and $[\phi_n^{ini}(x)]^T=[e^{ikx}+r_{ee}e^{-ikx}, r_{he}e^{ikx}]^T$ is that in 
the ballistic case with $r_{ee}$ and $r_{he}$ being the normal and the Andreev reflection coefficients 
at the NS interface from the electron branch, respectively. 
%The $2\times 2$ Green function $\hat{G}$ is obtained analytically.
By putting $x=x_i$, the equation in Eq.~(\ref{ls}) has the closed form for $\phi_n(x_i)$, 
which results in 
\begin{align}
&\phi_n(x)= \phi_n^{ini}(x)
%\nonumber\\
%&\phi_n^{ini}(x) + \hat{G}(E;x,x_i) V_i \hat{\sigma}_3
%\left[
%\hat{\sigma}_0-\hat{G}(E;x_i,x_i) V_i \hat{\sigma}_3\right]^{-1} \phi_n(x_i),\\
%\to &
%e^{ikx}\left[\begin{array}{c}
%1\\
%r_{he}\end{array}\right]\nonumber\\
 + \frac{1}{Y}\left[
 \begin{array}{c}
 -e^{ikx}e^{2ikx_i}z_i(z_i+i)X\\
 e^{ikx}(1-Y)r_{he}
\end{array}\right],\label{phin}
\end{align}
at $E=0$ for $x<x_i$,
where $Y=1+z_i^2X$, $X=1-r_{he}r_{eh}$, 
$z_i=V_i/\hbar v_F$, and 
$r_{eh}$ is the Andreev reflection coefficient of the NS interface from the hole 
branch.
We have already taken into account the absence of the normal reflection 
at the NS interface at $E=0$ (i.e., $r_{ee}=r_{hh}=0$). 
%The second term in Eq.~(\ref{phin}) represents effects 
%of scatterings by the single impurity on wave function. 
At $E=0$, the Andreev reflection coefficients $r_{he}=-ie^{-i\varphi}$ and 
$r_{eh}=ie^{i\varphi}$ satisfy an important relation 
%\begin{align}
$r_{eh}r_{he} =1$ 
%\label{condition}
%\end{align}
which eliminates the second term of Eq.~(\ref{phin}). Thus the zero-bias 
conductance quantization at 
 $G_Q=2e^2/h$ holds even in the presence of an impurity. 
The relation of $r_{eh}r_{he} =1$ is nothing but the condition for forming the MBS at $E=0$. 
To discuss whole effects of scatterings by many impurities, we need to solve the nonlinear 
quasiclassical Usadel equation~\cite{yt05r,yt04}. 
%The zero-bias resistance $R$ (inverse of the conductance) at $eV=0$ is given by
%the summation of the resistance of potential barrier at the NS interface $\tilde{R}_b$ and that of
%the normal metal $\tilde{R}_n$ as
%$R= \tilde{R}_b + \tilde{R}_n$. We note that $\tilde{R}_b$ and $\tilde{R}_b$ are the resistance 
%modified by the anomalous proximity effect. Thus they are different from the original 
%resistance in the normal state $R_b$ and $R_n$. The solution of the Usadel equation results in
%$\tilde{R}_b=G_Q^{-1}(1-\tanh(R_n G_Q))$ and $\tilde{R}_n=G_Q^{-1}\tanh(R_n G_Q)$. 
%As a consequence, $R$ remains at $G_Q^{-1}$ even in the diffusive NS junctions. 
%The original resistance of potential barrier ${R}_b$ does not appear in either $\tilde{R}_b$ 
%or $\tilde{R}_n$, which is also a results of Eq.~(\ref{condition}). 
The analytical expression of the zero-bias conductance also show $G_{NS}=G_Q$ (See Appendx C).
The diffusive normal metal is assumed in the Usadel equation. 
The validity of $G_\textrm{NS}=G_Q$ in the localization regime is confirmed in numerical calculation 
in Fig.~\ref{fig2}(a) and in Fig.~\ref{report_fig2}.

In summary, we have theoretically discussed the 
anomalous transport phenomena in NS and SNS junctions 
of nano wires in which the superconducting segment is topologically nontrivial
and the normal segment is disordered by random impurity potential.
The physics behind the anomalous transport can be understood in terms of 
the odd-frequency Cooper pairing. 
We conclude that Majorana fermions and odd-frequency Cooper pairs in solids 
are two sides of a same coin.

This work was supported by KAKENHI on Innovative Areas ``Topological Quantum Phenomena'' and
KAKENHI(22540355,22103005) from MEXT of Japan.

\appendix
\begin{widetext}
\section{Transformation of Hamiltonian}
 The starting Hamiltonian of this paper is equivalent to 
\begin{align}
H_{NW} =& \left[\begin{array}{cc}
\hat{h}_k & i\Delta\hat{\sigma}_2 e^{i\varphi}\\
-i\Delta\hat{\sigma}_2e^{-i\varphi} & -\hat{h}_{-k}^\ast
\end{array}\right],\\
h_{k}=&
\xi_k\hat{\sigma}_0 - V_{ex} \hat{\sigma}_3 - \lambda k \hat{\sigma}_2, \quad \xi_k= \frac{\hbar^2k^2}{2m}-\mu,
\end{align}
where $\mu$ is the chemical potential, $V_{ex}$ is the Zeeman potential due to 
external magnetic field, $\lambda k \hat{\sigma}_2$ represents the spin-orbit coupling, 
$\hat{\sigma}_0$ is the unit matrix in spin space, and $\hat{\sigma}_j$ for $j=1-3$ 
are the Pauli's matrices.
By applying following unitary transformation~\cite{sau}, the Hamiltonian is deformed as
\begin{align}
\check{D}^\dagger H_{NW}\check{D} =& 
\left[ \begin{array}{cccc}
\xi_k - A & 0 & \Delta_\lambda e^{i(\varphi+\pi/2)}& \Delta_{V}e^{i\varphi} \\
0 & \xi_k+A & -\Delta_{V}e^{i\varphi} & \Delta_\lambda e^{i(\varphi-\pi/2)} \\
\Delta_\lambda e^{-i(\varphi+\pi/2)}& -\Delta_V e^{-i\varphi}& -\xi_k + A & 0 \\
\Delta_Ve^{-i\varphi} &\Delta_\lambda e^{-i(\varphi-\pi/2)} & 0& -\xi_k -A 
\end{array}
\right], \\
 \Delta_\lambda=& \Delta \frac{\lambda k}{A}, \quad \Delta_V= \Delta \frac{V_{ex}}{A}, \quad A=\sqrt{V_{ex}^2+(\lambda k)^2}\\
\check{D}=&\left[\begin{array}{cc} \hat{U} & 0 \\ 0 & \hat{U} \end{array}\right],
\qquad 
\hat{U}=\left[\begin{array}{cc} \alpha &  i \, \textrm{sgn}(k) \beta \\  i\, \textrm{sgn}(k) \beta& \alpha
 \end{array}\right],\\
\alpha=& \sqrt{\frac{1}{2}\left(1+\frac{V_{ex}}{A}\right)}, \qquad
\beta= \sqrt{\frac{1}{2}\left(1-\frac{V_{ex}}{A}\right)}.
\end{align}
The topologically nontrivial phase is characterized by $V_{ex}>\sqrt{\mu^2+\Delta^2}$. In such case, 
only one dispersion remains at the fermi level for each Nambu space (i.e., $\xi_k+A > 0$). 
Therefore the Hamiltonian reduces
to $2\times 2$ Nambu space as
\begin{align}
\hat{H}_{NW2}=&\left[\begin{array}{cc} \xi_k-A & \Delta_\lambda e^{i(\varphi+\pi/2)}
\\ \Delta_\lambda e^{-i(\varphi+\pi/2)}  & -\xi_k+A \end{array}\right].
\end{align}
When we focus on the low-energy quasiparticle excitation, this is equivalent to 
the Hamiltonian describing the equal spin-triplet (spin less) $p_x$-wave superconductor
\begin{align}
\hat{H}_{p_x}=&\left[\begin{array}{cc} \xi_k & \Delta \frac{k}{k_F}e^{i\varphi} \\ \Delta\frac{k}{k_F}e^{-i\varphi}  & -\xi_k \end{array}\right]. \label{hpx}
\end{align}
Here we redefine $\mu+A \to \mu$ in the diagonal term and $\Delta_\lambda e^{i\pi/2} \to \Delta(k/k_F)$ 
in the off-diagonal term.
In previous papers~\cite{yt05r,ya06,ya07,tanaka07e,tanaka07,ya11}, we have studied the anomalous proximity effect starting from the 
Hamiltonian in Eq.~(\ref{hpx}).

\section{Analysis of $p_x$-wave superconductor}
\subsection{Green function and its representation}
The retarded Green functions are defined by the standard way
\begin{align}
\hat{G}(x,t;x',t') =& -i \Theta(t-t')
\left[ \begin{array}{cc}
\left\{ \psi(x,t), \psi^\dagger(x't') \right\} & \left\{ \psi(x,t), \psi(x't') \right\} \\
\left\{ \psi^\dagger(x,t), \psi^\dagger(x't') \right\} & \left\{ \psi^\dagger(x,t), \psi(x't') \right\} 
\end{array}\right],\label{Gdef1}\\
=&\left[ \begin{array}{cc}
G(x,t;x't') & F(x,t;x't') \\
\tilde{F}(x,t;x't') & \tilde{G}(x,t;x't')
\end{array}\right]\label{Gdef2},
\end{align}
where $\psi(x) (\psi^\dagger(x)) $ is the annihilation (creation) operator of a spin less electron.
In the case of spin-triplet superconductors, the electron operators are represented by the Bogoliubov transformation
\begin{align}
\left[\begin{array}{c} \psi(x) \\ \psi^\dagger(x) \end{array}\right]
=&
\sum_{\nu}\left[\begin{array}{cc} u_\nu(x) & v_\nu^\ast(x) \\ v_\nu(x) & u_\nu^\ast(x) 
\end{array}\right]
\left[\begin{array}{c} \gamma_\nu \\ \gamma_{-\nu}^\dagger \end{array}\right],\label{bogo}
%\\
%=&
%\sum_{\nu\neq 0}\left[\begin{array}{cc} u_\nu(x) & v_\nu^\ast(x) \\ v_\nu(x) & u_\nu^\ast(x) 
%\end{array}\right]
%\left[\begin{array}{c} \gamma_\nu \\ \gamma_{-\nu}^\dagger \end{array}\right]
%+\left[\begin{array}{c} \phi_0(x) \\ \phi_0^\dagger(x) \end{array}\right],\\
%\left[\begin{array}{c} \phi_0(x) \\ \phi_0^\dagger(x) \end{array}\right]
%=&\left[\begin{array}{cc} u_0(x) & v_0^\ast(x) \\ v_0(x) & u_0^\ast(x) 
%\end{array}\right]
%\left[\begin{array}{c} \gamma_0 \\ \gamma_{0}^\dagger \end{array}\right],\label{phi0}
\end{align}
where 
$\gamma_\nu$ is the  annihilation operators of Bogoliubov 
quasiparticle belonging to 
$E_\nu$. The wave function $u_\nu(x)$ and $v_\nu(x)$ 
are obtained by solving the 
Bogoliubov-de Gennes (BdG) equation.
The Green functions are expressed in spectral representation 
 as
\begin{align}
G(E;x,x')=&\sum_{\nu}\left[ 
\frac{u_\nu(x)u_\nu^\ast(x')}{E+i\delta - E_\nu} +\frac{v_\nu^\ast(x)v_\nu(x')}{E+i\delta + E_\nu} 
\right],\label{ge}\\
F(E;x,x')=&\sum_{\nu}\left[ 
\frac{u_\nu(x)v_\nu^\ast(x')}{E+i\delta - E_\nu} +\frac{v_\nu^\ast(x)u_\nu(x')}{E+i\delta + E_\nu} 
\right],\label{fe}
\end{align}
where $i\delta$ is a small imaginary part.

\subsection{Uniform superconductor}
The retarded Green function of a uniform spin less $p_x$-wave superconductor is calculated to be
\begin{align}
&\hat{G}(E;x,x')= \frac{-i \pi N_0}{2\Omega} \hat{\Phi}
\left[
\left(\begin{array}{cc} E+\Omega & \Delta s_x\\
\Delta s_x& E-\Omega \end{array}\right) e^{ik^+|x-x'|}
+
\left(\begin{array}{cc} E-\Omega & -\Delta s_x\\
-\Delta s_x& E+\Omega \end{array}\right) e^{-ik^-|x-x'|}
\right] \hat{\Phi}^\ast,\\
&\hat{G}(E;x,x')= \left(
\begin{array}{cc} G(E;x,x') & F(E;x,x') \\ 
\tilde{F}(E;x,x') & \tilde{G}(E;x,x') \end{array} \right), \\
&k^\pm= k\left(1\pm \frac{\Omega}{2\mu}\right), \quad \Omega=\sqrt{(E+i\delta)^2-\Delta^2},
\quad \hat{\Phi} =\textrm{diag}[e^{i\varphi/2},e^{-i\varphi/2}], \quad s_x=\textrm{sgn}(x-x'),
\end{align}
where $N_0$ is the density of states (DOS) at the fermi level in the normal state.
The Green functions are calculated as
\begin{align}
G(E;x,x')=& -i\frac{\pi N_0}{2} e^{ik\Omega|x-x'|/(2\mu)}
\left[ \frac{E}{\Omega}\cos k(x-x') +i \sin k|x-x'| \right],\\
%\tilde{G}(E,x,x')=& -i\frac{\pi N_0}{2} \left[ \frac{E}{\sqrt{E^2-\Delta^2}}\cos k(x-x') -i \sin k(x-x') \right],\\
-ie^{-i\varphi}F(E;x,x')=& -i\frac{\pi N_0}{2} e^{ik\Omega|x-x'|/(2\mu)} \frac{\Delta}{\Omega}\sin k(x-x').\label{funiform2}
\end{align}
From the normal Green functions, the local density of states (LDOS) is calculated to be
\begin{align}
N(E,x)=\frac{-1}{\pi}\textrm{Im} \textrm{Tr} \hat{G}(E,x,x) = N_0 \textrm{Re} \frac{E}{\Omega}.
\end{align}
The LDOS is an even function of $E$ because of the relation
\begin{align}
\sqrt{(E+i\delta)^2-\Delta^2}=&\left\{ 
\begin{array}{cc} 
\sqrt{E^2-\Delta^2} & \Delta <E \\
i\sqrt{\Delta^2-E^2} & 0 < |E| <\Delta  \\
-\sqrt{E^2-\Delta^2} & E <-\Delta 
\end{array}\right.
.
\end{align}
It is evident that there is no subgap state in uniform superconductor.
From the off-diagonal part, it is possible to check the pairing symmetry.
The anomalous Green function satisfies $F(x-x')=-F(x'-x)$, which indicates the odd-parity symmetry.
In addition, the real part of $-ie^{-i\varphi}F(E,x,x')$ is an even function of $E$, whereas the imaginary part of it 
is an odd function of $E$. This means that Cooper pairs have the even-frequency symmetry.

\subsection{Majorana surface bound state}
\begin{figure}[tbh]
\begin{center}
\includegraphics[width=5cm]{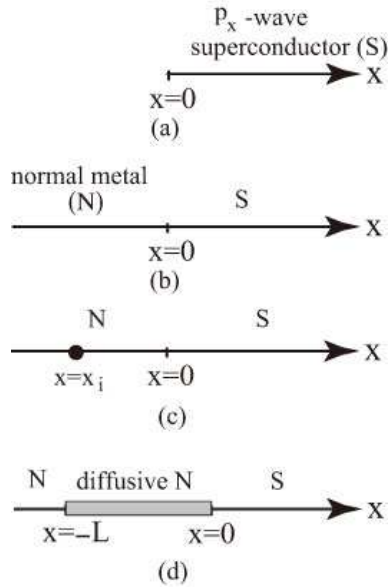}
\end{center}
\caption{System under consideration.  (a): a semi-infinite $p_x$-wave superconductor.
(b): a clean normal-metal/superconductor (NS) junction of $p_x$-wave symmetry.
(c): an impurity is introduced in the normal metal. 
(d): a diffusive normal metal is introduced in the NS junction.
}
\label{report_fig1}
\end{figure}
Next we consider a semi-finite $p_x$-wave superconductor which occupies $x>0$ as shown in Fig.~1(a).
By solving the Bogoliubov-de Gennes equation, the wave function for subgap state is expressed by
\begin{align}
\Psi_S(x)=&
A\left[ \begin{array}{c} E + i\tilde{\Omega} \\ \Delta e^{-i\varphi} \end{array} \right]
e^{ik x}e^{-x/2\xi_0} + 
B\left[ \begin{array}{c} E - i\tilde{\Omega} \\ -\Delta e^{-i\varphi} \end{array} \right]
e^{-ikx}e^{-x/2\xi_0},
\end{align}
where $\tilde{\Omega}=\sqrt{\Delta^2-E^2}$, $A$ and $B$ are constant.
From the boundary condition at $x=0$ (i.e., $\Psi_S(x=0)=0$), we find that a subgap state 
exists at $E=0$ and that the wave function of it becomes
\begin{align}
\Psi_S(x)=& 
\left[ \begin{array}{c} u_0(x) \\ v_0(x) \end{array} \right]
=
C(x)
\left[ \begin{array}{c} e^{i\pi/4} e^{i\varphi/2} \\ e^{-i\pi/4} e^{-i\varphi/2} \end{array} \right],
\quad
C(x) =\sqrt{\frac{2}{\xi_0}} e^{-x/2\xi_0}\sin (kx).\label{majo1}
\end{align}
We note that two components in the wave function satisfy an important relation
\begin{align}
u_0(x)=v_0^\ast(x). \label{majo2}
\end{align}

The BdG transformation reads,
\begin{align}
\left[\begin{array}{c} \psi(x) \\ \psi^\dagger(x) \end{array}\right]
%=&
%\sum_{\nu}\left[\begin{array}{cc} u_\nu(x) & v_\nu^\ast(x) \\ v_\nu(x) & u_\nu^\ast(x) 
%\end{array}\right]
%\left[\begin{array}{c} \gamma_\nu \\ \gamma_{-\nu}^\dagger \end{array}\right]\\
=&
\sum_{\nu\neq 0}\left[\begin{array}{cc} u_\nu(x) & v_\nu^\ast(x) \\ v_\nu(x) & u_\nu^\ast(x) 
\end{array}\right]
\left[\begin{array}{c} \gamma_\nu \\ \gamma_{-\nu}^\dagger \end{array}\right]
+\left[\begin{array}{c} \phi_0(x) \\ \phi_0^\dagger(x) \end{array}\right],\\
\left[\begin{array}{c} \phi_0(x) \\ \phi_0^\dagger(x) \end{array}\right]
=&\left[\begin{array}{cc} u_0(x) & v_0^\ast(x) \\ v_0(x) & u_0^\ast(x) 
\end{array}\right]
\left[\begin{array}{c} \gamma_0 \\ \gamma_{0}^\dagger \end{array}\right],\label{phi0}
\end{align}
where 
$\gamma_0$ is the annihilation operator of the bound state.
%The normal and the anomalous Green functions are defined as
%\begin{align}
%g(xt,x't')= -i \Theta(t-t')\left\langle
%\left\{ \psi(x,t), \psi^\dagger(x',t')\right\} \right\rangle, \\
%f(xt,x't')= -i \Theta(t-t')\left\langle
%\left\{ \psi(x,t), \psi(x',t')\right\} \right\rangle, 
%\end{align}
%In the spectrum representation they become
%\begin{align}
%g(E;x,x')=&\sum_{\nu}\left[ 
%\frac{u_\nu(x)u_\nu^\ast(x')}{E+i\delta - E_\nu} +\frac{v_\nu^\ast(x)v_\nu(x')}{E+i\delta + E_\nu} 
%\right],\\
%f(E;x,x')=&\sum_{\nu}\left[ 
%\frac{u_\nu(x)v_\nu^\ast(x')}{E+i\delta - E_\nu} +\frac{v_\nu^\ast(x)u_\nu(x')}{E+i\delta + E_\nu} 
%\right],
%\end{align}
%where $i\delta$ is the small imaginary part.
%
%The relation of electron operators and Bogoliubov operators for the surface bound state 
%should be considered separately as 
%\begin{align}
%\left[\begin{array}{c} \psi_0(x) \\ \psi^\dagger_0(x) \end{array}\right]
%=\left[\begin{array}{cc} u_0(x) & v_0^\ast(x) \\ v_0(x) & u_0^\ast(x) 
%\end{array}\right]
%\left[\begin{array}{c} \gamma_0 \\ \gamma_{0}^\dagger \end{array}\right]
%\end{align}
Together with Eqs.~(\ref{majo1}) and (\ref{majo2}), we find
\begin{align}
\phi_0(x) =& e^{i\pi/4} e^{i\varphi/2} \gamma(x),\quad
\phi_0^\dagger (x) =  e^{-i\pi/4} e^{-i\varphi/2} \gamma(x),\label{majo3}\\
\gamma(x)=&C(x) (\gamma_0 + \gamma_0^\dagger).
\end{align}
The fermion operator $\gamma(x)$ satisfies the Majorana relation $\gamma(x)=\gamma^\dagger(x)$.
When we focus on $|E| \ll \Delta$,
the contributions from $\nu=0$ become dominant in Eqs.~(\ref{ge}) and (\ref{fe}).
Near $E=0$, the normal and the anomalous Green functions satisfy a relation 
\begin{align}
G(E;x,x')=-ie^{-i\varphi}F(E;x,x'), \label{gf2}
\end{align}
because they are calculated from Eq.~(\ref{phi0}) as
\begin{align}
G(E;x,x')=\frac{2C(x)C(x')}{E+i\delta},\quad F(E;x,x')=\frac{2C(x)C(x')}{E+i\delta}ie^{i\varphi}.
%-i \int_0^{\infty} d(t-t')e^{iE(t-t')} \left\langle \left\{\psi_0(x), \psi_0^\dagger(x')\right\}\right\rangle,\\
%f(E;x,x')=-i \int_0^{\infty} d(t-t') e^{iE(t-t')} \left\langle \left\{\psi_0(x), \psi_0(x')\right\}\right\rangle.
\end{align}
 Eq.~(\ref{gf2}) directly relates Majorana fermions and odd-frequency Cooper pairs.
%The imaginary part of $G(E;x,x)$ gives the local density of the Andreev bound state and 
%must be an even function of $E$. 
When we consider $x=x'$, $F(E,x,x)$ represents the pairing function of $s$-wave symmetry.
The real part of $-ie^{-i\varphi}F(E;x,x)$ is an odd-function of $E$ and
the imaginary part of it is an even function of $E$. This indicates that Cooper pairs have the 
odd-frequency symmetry. It is possible to confirm Eq.~(\ref{gf2}) in NS junctions as discussed below.

\subsection{NS junction of $p_x$ superconductor}
It is possible to calculate the Green function of a junction which consists of a 
normal metal $(x<0)$ and a $p_x$-wave superconductor ($x>0$) in one dimension as shown in Fig.~1(b).
In the superconductor, the retarded Green function becomes~\cite{ya04-2}
\begin{align}
&\hat{G}_{ss}(E;x,x')=\hat{\Phi}\-i\frac{N_0\pi}{2}\frac{E}{\Omega} \nonumber \\
\times&
\left[
\left(\begin{array}{cc}u^2 & uvs_x \\ uv s_x& v^2\end{array} \right)e^{ik^+|x-x'|}+
\left(\begin{array}{cc}v^2 & -uvs_x \\ -uvs_x & u^2\end{array} \right)e^{-ik^-|x-x'|}\right. \nonumber\\
&+\left(\begin{array}{cc}-uv & v^2 \\ u^2 & -uv \end{array} \right)
e^{-ik^-x+ik^+x'} r^{he}_{ss} 
+\left(\begin{array}{cc}uv & u^2 \\ v^2 & uv \end{array} \right)
e^{ik^+x-ik^-x')} r^{eh}_{ss} \nonumber\\
& +\left(\begin{array}{cc}u^2 & -uv \\ uv & -v^2 \end{array} \right)
e^{ik^+(x+x')}r^{ee}_{ss} +
\left. +\left(\begin{array}{cc}-v^2 & -uv \\ uv & u^2 \end{array} \right)
e^{-ik^-(x+x')} r^{hh}_{ss} \right] \hat{\Phi}^\ast, \label{gss}
\end{align}
\begin{align}
&\hat{G}_{ss}(E;x,x')= \left(
\begin{array}{cc} G_{ss}(E;x,x') & F_{ss}(E;x,x') \\ 
\tilde{F}_{ss}(E;x,x') & \tilde{G}_{ss}(E;x,x') \end{array} \right),\quad
\hat{\Phi}=\textrm{diag}(e^{i\varphi/2},e^{-i\varphi/2}),\\
u(v)=& \sqrt{\frac{1}{2}\left( 1+(-)\frac{\Omega}{E}\right) }, \quad
\Omega=\sqrt{(E+i\delta)^2-\Delta^2}, \quad k^{\pm}=k\left(1 \pm \frac{\Omega}{2\mu}\right), \quad
s_x=\textrm{sgn}(x-x')
\end{align}
for $x, x' >0$.
%Here $\varphi$ is the superconducting phase, 
%and $k$ is the Fermi wave number in one-dimension.
The subscript $ss$ in the Green function $\hat{G}_{ss}(E;x,x')$ means that both $x>0$ and $x'>0$ indicate 
places in the superconductor.
The normal and Andreev reflection coefficients are given by
\begin{align}
r^{he}_{ss}=& \frac{uv}{\Xi}(2-|t_n|^2) = -r^{eh}_{ss},\qquad
r^{ee}_{ss}=\frac{r}{\Xi}(u^2-v^2), \quad r^{hh}_{ss}=\frac{r^\ast}{\Xi}(u^2-v^2),\\
\Xi =& 1-|t_n|^2 v^2, \quad t_n=\frac{k}{k+iz_0}, \quad r_n=\frac{-iz_0}{k+iz_0}, \quad z_0 =V_0/\hbar v_F
\end{align}
where $t_n$ and $r_n$ are the normal transmission coefficients 
due to the potential barrier at the interface described by $V_0\delta(x)$.
When we focus on the subgap energy in the tunneling limit (i.e., $|E|\ll \Delta$ and $|t_n|\ll 1$), we 
find, 
\begin{align}
G_{ss}(E;x,x)
%=\tilde{G}_{ss}(E;x,x)
=-ie^{-i\varphi}F_{ss}(E;x,x)\approx 
\pi N_0 \frac{\Delta}{E+i\Delta|t_n|^2/2}e^{-x/\xi_0} \sin^2(kx).
\end{align}
The imaginary part of $G_{ss}(E;x,x)$ gives the local density of the Andreev bound state and 
must be an even function of $E$. Therefore the real part of $-ie^{-i\varphi}F_{ss}(E;x,x)$ is an odd-function of $E$ and
the imaginary part of it is an even function of $E$, which indicates the 
odd-frequency symmetry. The condition in Eq.~(\ref{majo2}) leads to the Majorana relation in operators at 
Eq.~(\ref{majo3}).  Eq.~(\ref{majo3}) results in Eq.~(\ref{gf2}). Then Eq.~(\ref{gf2}) guarantees the
odd-frequency symmetry of Cooper pairs.  
The orbital part of Cooper pairs is $s$-wave symmetry because $-ie^{-i\varphi}F_{ss}$ is calculated at $x=x'$.

\section{Perfect transmission at $E=0$}
In the normal metal of NS junction ($x,x'<0$), the Green function is given by~\cite{ya04-2}
\begin{align}
&\hat{G}_{nn}(E;x,x') =-i \frac{\pi N_0}{2}
\left[
\begin{array}{cc}
e^{-ik|x-x'|} + e^{-ik(x+x')} r_{nn}^{ee} & e^{-ikx+ikx'} r_{nn}^{eh} \\
e^{ikx-ikx'} r_{nn}^{he} & e^{ik|x-x'|} + e^{ik(x+x')} r_{nn}^{hh}
\end{array}\right],\\
&r_{nn}^{he}= \frac{|t_n|^2 e^{-i\varphi} uv}{\Xi}, \quad r_{nn}^{eh}= -\frac{|t_n|^2 e^{i\varphi} uv}{\Xi},
\quad r_{nn}^{ee}= \frac{r_n}{\Xi}, \quad r_{nn}^{hh}= \frac{r_n^\ast}{\Xi},\\
&\hat{G}_{nn}(E;x,x')= \left(
\begin{array}{cc} G_{nn}(E;x,x') & F_{nn}(E;x,x') \\ 
\tilde{F}_{nn}(E;x,x') & \tilde{G}_{nn}(E;x,x') \end{array} \right).\\
\end{align}
The subscript $nn$ in the Green function $\hat{G}_{nn}(E;x,x')$ means that both $x<0$ and $x'<0$ indicate 
places in the normal metal.
We also confirmed the relation between the Green function in the normal metal,
\begin{align}
G_{nn}(E;x,x)=-ie^{-i\varphi}F_{nn}(E;x,x)\approx  \pi N_0 \frac{\Delta|t_n|^2}{E+ i\Delta|t_n|^2/2},
\end{align}
for $|E|\ll \Delta$ and $|t_n|^2 \ll 1$.
As we discussed in Eq.~(\ref{gf2}), this suggests the presence of Majorana fermions and 
odd-frequency Cooper pairs in the normal metal. 

It is possible to calculate exactly the wave function where a single impurity $V_i\delta(x-x_i)$ 
exist in the normal metal as shown in Fig.~1(c) by using the Lippmann-Schwinger equation,
\begin{align}
\phi_n(x)=&\phi_n^{ini}(x) + \hat{G}_{nn}(E;x,x_i) V_i \hat{\sigma}_3 \phi_n(x_i),\\
\phi_n^{ini}(x)=&\left[\begin{array}{c}
e^{ikx} + e^{-ikx}r_{nn}^{ee} \\
e^{ikx}r_{nn}^{he}\end{array}\right]
\end{align}
where $\phi_n^{ini}(x)$ is the wave function in the ballistic case and $\phi_n(x)$ is that in the presence of 
the impurity.
By solving this equation at $x=x_i$, we obtain 
\begin{align}
\phi_n(x_i) = \left[
1-\hat{G}_{nn}(E;x_i,x_i) V_i \hat{\sigma}_3\right]^{-1} \phi_n^{ini}(x_i).
\end{align}
The wave function for $x<x_i$ in the presence of the single impurity is expressed by
\begin{align}
\phi_n(x)=&\phi_n^{ini}(x) + \hat{G}_{nn}(E;x,x_i) V_i \hat{\sigma}_3
\left[
1-\hat{G}_{nn}(E;x_i,x_i) V_i \hat{\sigma}_3\right]^{-1} \phi_n^{ini}(x_i),\\
=&\phi_n^{ini}(x)
+ \frac{1}{Y}\left[\begin{array}{c}
e^{-ikx}\left\{ -iz_i(B_1^2- e^{2ikx_i}r_{nn}^{he}r_{nn}^{eh}\right\} 
-z_i^2\left\{B_1B_2-r_{nn}^{he}r_{nn}^{eh}\right\}e^{ikx_i}B_1 \\
e^{ikx} (1-Y) r_{nn}^{he} 
\end{array}\right],\\
Y=&1+z_i(e^{i2kx_i} r_{nn}^{hh} -e^{-i2kx_i} r_{nn}^{ee})
+z_i^2(1- r_{nn}^{he}r_{nn}^{eh}+ e^{i2kx_i} r_{nn}^{hh} + e^{-i2kx_i} r_{nn}^{ee}),\\
B_1=&e^{ikx_i}+e^{-ikx_i}r_{nn}^{ee},\quad  B_2=e^{-ikx_i}+e^{ikx_i}r_{nn}^{hh}, \quad z_i=V_i/\hbar v.
\end{align}
At $E=0$, the reflection coefficients become
\begin{align}
r_{nn}^{eh} =ie^{i\varphi},\quad r_{nn}^{he} =-ie^{-i\varphi}, \quad r_{nn}^{ee} =r_{nn}^{hh}=0.
\end{align}
These relations immediately lead to
\begin{align}
B_1^2- e^{2ikx_i}r_{nn}^{he}r_{nn}^{eh}=B_1B_2-r_{nn}^{he}r_{nn}^{eh}=0, \quad
Y=1.
\end{align}
Therefore we find that the wave function in the presence of the single impurity at $x=x_i$ 
remains unchanged from the original one
\begin{align}
\phi_n(x)=
\left[\begin{array}{c}
e^{ikx} \\
0 
\end{array}\right] +\left[\begin{array}{c}
0 \\
-ie^{-i\varphi} e^{ikx} 
\end{array}\right]. \label{wave2}
\end{align}
The first term represents the incoming wave at the electron branch. The second term expresses
 the 
outgoing wave in the hole branch.
The Andreev reflection is perfect and the normal reflection is absent even in the 
presence of the single impurity in the normal metal. 
With using the Blonder-Tinkham-Klapwijk formula, 
\begin{align}
G_{NS}=\frac{e^2}{h}\left[ 1 -|r_{nn}^{ee}|^2 + |r_{nn}^{he}|^2\right],
\end{align}
the zero-bias conductance of the NS junction
remains unchanged from $G_{NS}=2e^2/h$ independent of the impurity scattering. 
%The conclusion remains unchanged even when
%we introduce the second impurity. 

 On the way to the conclusion, we derive a relation 
\begin{align}
r_{nn}^{eh} r_{nn}^{he}=1, \label{a1a2}
\end{align}
 at $E=0$. 
This plays an important role in the resonant transmission of a quasiparticle 
in a normal metal.
For comparison, the Andreev reflection coefficients of the $s$-wave transparent NS junction 
becomes 
\begin{align}
r_{nn}^{he}=-ie^{-i\varphi}, \quad r_{nn}^{eh}=-ie^{i\varphi},
\end{align}
at $E=0$. However, they gives a relation $r_{nn}^{eh} r_{nn}^{he}=-1$.
In this case, a quasiparticle is scattered by the impurity and the conductance 
decreases from $G_{NS}=G_Q$.
The relation in Eq.~(\ref{a1a2}) is equivalent to the necessary condition for
 the formation of Andreev (Majorana) bound states at $E=0$.

\section{Analysis of quasiclassical Usadel Equation}
In this section, we consider a diffusive normal metal is attached to $p_x$-wave
superconductor as shown in Fig.~1(d).
At first, we define the quasicalssical Green functions in terms of Gor'kov Green functions.
In the mixed representation, Gor'kov Green functions become
\begin{align}
{G}(x,t;x',t')=&{G}( x_c,x-x',t_c,t-t')=
\int \frac{d\epsilon}{2\pi} \int \frac{dk}{2\pi}
G(x_c,k,t_c,\epsilon)e^{ik(x-x')-i\epsilon(t-t')},\\
{F}(x,t;x',t')=&{F}( x_c,x-x',t_c,t-t')=
\int \frac{d\epsilon}{2\pi} \int \frac{dk}{2\pi}
F(x_c,k,t_c,\epsilon)e^{ik(x-x')-i\epsilon(t-t')},\\
x_c=&\frac{x+x'}{2}, \quad t_c=\frac{t+t'}{2}.
\end{align}
When we consider static state, the Green functions are independent of $t_c$.
With replacing $x_c$ by $x$, the quasiclassical Green functions are defined as
\begin{align}  
g(x,k,\epsilon) =& \frac{i}{\pi} \int d\xi_k G(x,k,\epsilon) - \frac{i}{\pi} \int d\xi_k \frac{\mathcal{P}}{\xi_k},\\
f(x,k,\epsilon) =& \frac{i}{\pi} \int d\xi_k F(x,k,\epsilon).
\end{align}
They obey so called Eilengerger equation. In what follows, we fix the phase of the superconductor $\varphi$ at 0.
When the normal metal is in the dirty limit, $g(x,k,\epsilon)$ and $f(x,k,\epsilon)$ 
are isotropic in momentum space.
Since they satisfy 
 the normalization condition $g^2(x,\epsilon)+f^2(x,\epsilon)=1$, it is possible to 
 apply a parameterization: $g(x,\epsilon)=\cos[\theta(x,\epsilon)]$ and $f(x,\epsilon)=\sin[\theta(x,\epsilon)]$. 
The function $\theta(x,\epsilon)$ obeys the Usadel equation in the diffusive normal metal
\begin{align}
D\frac{\partial^2 \theta(x,\epsilon)}{\partial x^2} + 2i \epsilon \sin[\theta(x,\epsilon)]=0,\label{usadel}
\end{align}
where $D$ is the diffusion constant in the dirty normal metal.

In what follows, we consider NS junction shown in Fig. 1(d), where a dirty normal metal is introduced between 
a clean normal lead wire ($x<-L$) and a $p_x$-wave superconductor ($x>0$).
The boundary condition for $\theta(x,\epsilon)$ are given by~\cite{yt05r}
\begin{align}
\theta(x=-L,\epsilon)=&0,\\
\frac{L}{R_N} \left.\frac{\partial \theta(x,\epsilon)}{\partial x} \right|_{x=0}
=& \frac{2}{R_B} \frac{ f_S\cos\theta_0 - g_S \sin\theta_0 }
{ 2-|t_n|^2+|t_n|^2(f_S\sin\theta_0+g_S\cos\theta_0)},
\end{align}
with
\begin{align}
\theta_0=& \theta(x=0,\epsilon), \quad g_S = \frac{ g_+ + g_-}{ 1 + g_+  g_- + f_{+} f_{-} }, \quad
f_S=i\frac{f_+g_--f_-g_+}{1+g_+g_-+f_+f_-},\\
R_B=&\left[ G_Q |t_n|^2 \right]^{-1}, \quad G_Q=\frac{2e^2}{h}.
\end{align}
The parameter $R_N$ and $R_B$ are the normal resistance of the dirty normal metal and that due to the potential 
barrier at the NS interface, respectively.
The information of the pairing symmetry of superconductor is embedded in the surface Green function $g_S$ and $f_S$.
The total resistance of the junction $R$ is represented by~\cite{yt05r}
\begin{align}
R=& \tilde{R}_B + \tilde{R}_N,\\
\tilde{R}_B =&
\frac{1}{2}\frac{C_0}{|(2-|t_n|^2) +|t_n|^2( \cos\theta_0 g_S + \sin\theta_0 f_S)|^2},\\
C_0=& |t_n|^2 (1+|\cos\theta_0|^2+|\sin\theta_0|^2)(1+|g_S|^2+|f_S|^2) \nonumber\\
+4(2-|t_n|^2)&\left[ \textrm{Re}(g_S)\textrm{Re}(\cos\theta_0) +\textrm{Re}(f_S)\textrm{Re}(\sin\theta_0) \right]
+4|t_n|^2 \textrm{Im}(\cos\theta_0\sin^\ast\theta_0) \textrm{Im}(g^\ast_S f_S),\\
\tilde{R}_N=& 
\frac{R_N}{L} \int_{-L}^{0} \frac{2dx}{1+|\cos\theta(x,\epsilon)|^2+|\sin\theta(x,\epsilon)|^2}.
\end{align}
The resistance $\tilde{R}_B$ and $\tilde{R}_N$ are not equal to their normal one's $R_B$ and $R_N$.
They are modified by the proximity effect.

In the case of the $p_x$-wave superconductor, following relations hold
\begin{align}
g_+=g_-=\frac{\epsilon}{\sqrt{(\epsilon+i0^+)^2-\Delta^2}}, \quad
f_+=-f_-=\frac{i\Delta}{\sqrt{(\epsilon+i0^+)^2-\Delta^2}}.
\end{align} 
The $p_x$-wave symmetry of superconductor is represented by the relation $f_+=-f_-$.
At the surface of $p_{x}$-wave superconductor, purely odd-frequency pairing state 
exist due to the formation of Andreev bound state as discussed in Sec.~2.3. 
\begin{figure}[tbh]
\begin{center}
\includegraphics[width=7cm]{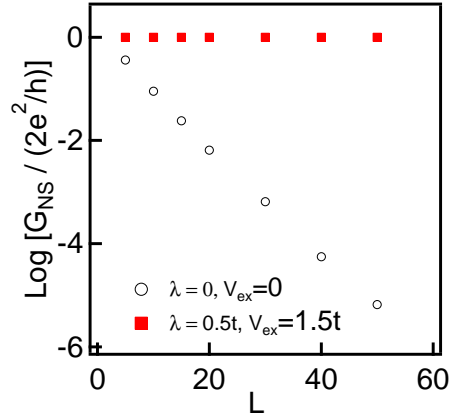}
\end{center}
\caption{ Conductance are plotted as a function of $L$ in NS junctions of nano wire.
}
\label{report_fig2}
\end{figure}

The Usadel equation can be solved analytically at $\epsilon=0$. 
Under the boundary condition $(L/R_N)(\partial \theta/\partial x)|_{x=0}=iG_Q$, we obtain
\begin{align}
\theta(x,\epsilon=0)=&iR_NG_Q \frac{x+L}{L},\\
g(x)=&\cos\theta(x,0)=\cosh\left(R_NG_Q\frac{x+L}{L}\right), \\
f(x)=&\sin\theta(x,0)=i\sinh\left(R_NG_Q\frac{x+L}{L}\right),
\end{align}
at $\epsilon=0$. The pairing function $f(x)$ represents the spin-triplet $s$-wave odd-frequency pair.
Indeed, $f(x)$ is purely imaginary number at $\epsilon=0$.
Finally we obtain the zero-bias resistance 
\begin{align}
R=&\tilde{R}_B+\tilde{R}_N=G_Q^{-1},\\
\tilde{R}_B=& \frac{1}{G_Q}\left[1+if(0)/g(0)\right]= \frac{1}{G_Q} [1-\tanh(G_QR_N)], \\
\tilde{R}_N=& \frac{1}{G_Q}\left[-if(0)/g(0)\right] =\frac{1}{G_Q}\tanh(G_QR_N).
\end{align}
The total resistance at the zero-bias voltage is independent of $R_N$ and $R_B$, and remains unchanged from $R=G_Q^{-1}$. 
It is worth to consider the physical meaning of the 
resulting resistances $\tilde{R}_B$ and $\tilde{R}_N$. 
$\tilde{R}_B$ is the resistance at the interface which
decreases from $G_Q^{-1}$ with the increase of $R_{N}$.
In other words, the interface resistance decreases 
with the increase of the amplitude of odd-frequency pair $f(0)$. 
$\tilde{R}_N$ is the resistance of the dirty normal metal.
In the limit of weak proximity effect $R_{N}G_Q \ll 1$, the amplitude of 
odd-frequency pairs becomes small. 
In such limit, we find $\tilde{R}_{N}=R_N$. 
On the other hand for $R_{N}G_Q\gg 1$, $-if(0)/g(0)$ goes to unity and $\tilde{R}_N$ approaches to 
$G_Q^{-1}$. 
Thus the odd-frequency pairs play a crucial role in the relation 
of $R=G_Q^{-1}$.

The diffusive transport is assumed in the Usadel equation. It is possible to check the validity 
of $G_\textrm{NS}=G_Q$ when the normal segment is in the localization regime.
In Fig.~\ref{report_fig2}, we plot the zero-bias conductance in NS junctions of nano wire as a function of 
the length of the disordered segment $L$.
When the nano wire is non topological at $\lambda=V_{ex}=0$, the zero-bias conductance 
decreases exponentially with $L$ due to the localization. 
On the other hand, the conductance for topological nano wire junctions at $\lambda=0.5t$ and $V_{ex}=1.5t$
remains unchanged from $G_Q$ even in the localization regime.
\end{widetext}


\begin{thebibliography}{99}
\bibitem{localization}
The normal conductance $g_{\textrm{N}}$ in units of $2e^{2}/h$ 
well obeys a relation of
$ \ln(g_{\textrm{N}})  \propto -L/\xi_{\text{AL}}$ with $\xi_{\text{AL}}$ 
with being the localization length. 
We estimate that $\xi_{\text{AL}}$ for the topological ($V_{ex}=\lambda=0$) and 
the non-topological ($V_{ex}=1.5t, \lambda=0.5t$) nano wires are about 11 and 8.5 lattice constants, respectively.
\end{thebibliography}
\end{document}